# AdSplit: Separating smartphone advertising from applications

Shashi Shekhar
*shashi.shekhar@rice.edu*

Michael Dietz
*mdietz@rice.edu*

Dan S. Wallach
*dwallach@rice.edu*

## Abstract

A wide variety of smartphone applications today rely on third-party advertising services, which provide libraries that are linked into the hosting application. This situation is undesirable for both the application author and the advertiser. Advertising libraries require additional permissions, resulting in additional permission requests to users. Likewise, a malicious application could simulate the behavior of the advertising library, forging the user's interaction and effectively stealing money from the advertiser. This paper describes AdSplit, where we extended Android to allow an application and its advertising to run as separate processes, under separate user-ids, eliminating the need for applications to request permissions on behalf of their advertising libraries. We also leverage mechanisms from Quire to allow the remote server to validate the authenticity of client-side behavior. In this paper, we quantify the degree of permission bloat caused by advertising, with a study of thousands of downloaded apps. AdSplit automatically recompiles apps to extract their ad services, and we measure minimal runtime overhead. We also observe that most ad libraries just embed an HTML widget within and describe how AdSplit can be designed with this in mind to avoid any need for ads to have native code.

## 1 Introduction

The smartphone and tablet markets are growing in leaps and bounds, helped in no small part by the availability of specialized third-party applications ("apps"). Whether on the iPhone or Android platforms, apps often come in two flavors: a free version, with embedded advertising, and a pay version without. Both models have been successful in the marketplace. To pick one example, the popular Angry Birds game at one point brought in roughly equal revenue from paid downloads on Apple iOS devices and from advertising-supported free downloads on Android devices [10]. They now offer advertising-supported free downloads on both platforms.

We cannot predict whether free or paid apps will dominate in the years to come, but advertising-supported applications will certainly remain prominent. Already, a cottage industry of companies offer advertising services for smartphone application developers.

Today, these services are simply pre-compiled code libraries, linked and shipped together with the application. This means that a remote advertising server has no way to validate a request it receives from a user legitimately clicking on an advertisement. A malicious application could easily forge these messages, generating revenue for its developer while hiding the advertisement displays in their entirety. To create a clear trust boundary, advertisers would benefit from running separately from their host applications.

In Android, applications must request permission at install time for any sensitive privileges they wish to exercise. Such privileges include access to the Internet, access to coarse or fine location information, or even access to see what other apps are installed on the phone. Advertisers want this information to better profile users and thus target ads at them; in return, advertisers may pay more money to their hosting applications' developers. Consequently, many applications which require no particular permissions, by themselves, suffer *permission bloat*—being forced to request the privileges required by their advertising libraries in addition to any of their own needed privileges. Since users might be scared away by detailed permission requests, application developers would also benefit if ads could be hosted in separate applications, which might then make their own privilege requests.

And, of course, separating applications from their advertisements creates better fault isolation. If the ad system fails or runs slowly, the host application should be able to carry on without inconveniencing the user. Addressing these needs requires developing a suitable soft-ware architecture, with OS assistance to make it robust.

The rest of the paper is organized as follows: first in Section 2 we present a survey of thousands of Android applications, and estimate the degree of permission bloat caused by advertisement libraries. Then in Section 3.1 we discuss advertisement security on web and the challenges for securing advertisements on smartphones. Section 3 describes the design of AdSplit, Section 4 describes our Android-based implementation, and Section 5 quantifies its performance. Section 6 provides details about a simple binary rewriter to adapt legacy apps to use our system. Section 7 considers how we might eliminate native code libraries for advertisements and go with a more web-like architecture. Finally, Section 8 discusses a variety of policy issues.

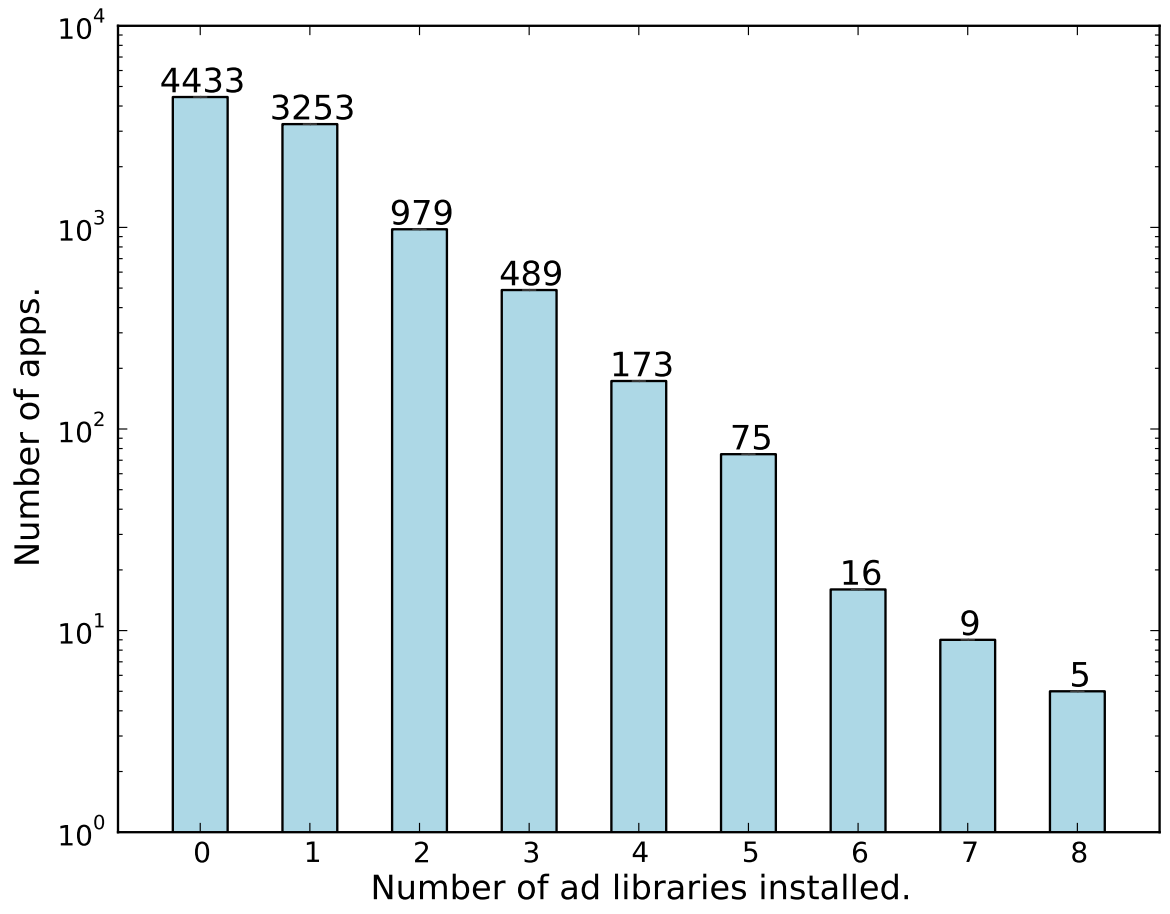

Figure 1: Number of apps with ad libraries installed.

## 2 App analysis

The need to monetize freely distributed smartphone applications has given rise to many different ad provider networks and libraries. The companies competing for business in the mobile ad world range from established web ad providers like Google's AdMob to a variety of dedicated smartphone advertising firms.

With so many options for serving mobile ads, many app developers choose to include multiple ad libraries. Additionally, there is a new trend of *advertisement aggregators* that allow an application developer to include multiple ad libraries with their app and have the aggregator choose which ads to display in order to maximize profits for the developer.

While we're not particularly interested in advertising market share, we want to understand how these ad libraries behave. What permissions do they require? And how many apps would be operating with fewer permissions, if only their advertisement systems didn't require them? To address these questions, we downloaded approximately 10,000 free apps from the Android Market and the Amazon App Store and built tools to analyze them.

**How many ad libraries?** Fig 1 shows the distribution of number of advertisement libraries used by apps in our sample. Of the apps that use advertisements, about 35% include two or more advertising libraries.

**Permissions required.** We found that some ad libraries need more permissions than those mentioned in the documentation, also the set of permissions may change with the version of the ad library. Table 1 shows some of the required and optional permission sets for a number of popular Android ad libraries. The permissions listed as optional are not required to use the ad library but may be requested in order to improve the quality of advertisements; for example, some ad libraries will use location information to customize ads. A developer using such a library has the choice of including location-targeted ads or not. Presumably, better targeted ads will bring greater revenue to the application developer.

| Ad Library | Internet | NetworkState | ReadPhoneState | WriteExternalStorage | CoarseLocation | CallPhone |
|---|---|---|---|---|---|---|
| AdMob [22] | ✓ | ✓ | | | ○ | |
| Greystripe [24] | ✓ | ✓ | ✓ | | | |
| Millennial Media [34] | ✓ | ✓ | ✓ | ✓ | | |
| InMobi [28] | ✓ | ○ | | | ○ | ○ |
| MobClix [36] | ✓ | ○ | ✓ | | | |
| TapJoy [48] | ✓ | ✓ | ✓ | ✓ | | |
| JumpTap [31] | ✓ | ✓ | ✓ | | ○ | |

✓(required), ○ (optional)

Table 1: Different advertising libraries require different permissions.

**Permission bloat.** In Android, an application requests a set of permissions at the time it's installed. Those permissions must suffice for all of the app's needs and for the needs of its advertising library. We decided to mea-

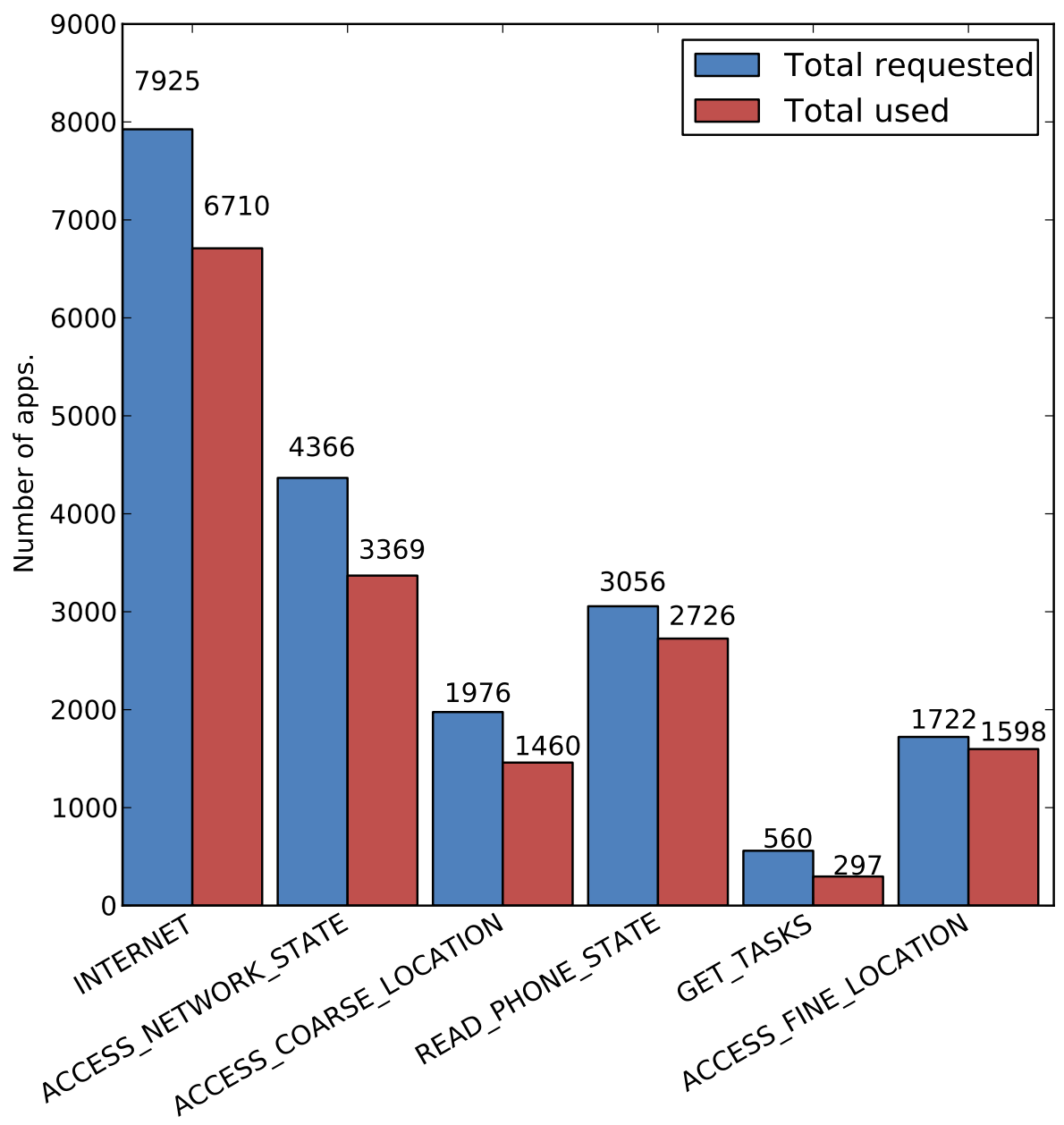

Figure 2: Distribution of types of permissions reduced when advertisements are separated from applications.

sure how many of the permissions requested are used *exclusively* by the advertising library (i.e., if the advertising library were removed, the permission would be unnecessary).

This analysis required decompiling our apps into dex format [3] using the android-apktool [23]. For each app, we then extracted a list of all API calls made. Since advertising libraries have package names that are easy to distinguish, it's straightforward to separate their API calls from the main application. To map the list of API calls to the necessary permissions, we use the data gathered by Felt et. al [18]. This allows us to compute the minimal set of permissions required by an application, with and without its advertisement libraries. We then compare this against the formal list of permissions that each app requests from the system.

There may be cases where an app speculatively attempts to use an API call that requires a permission that was never granted, or there may be dead code that exercises a permission, but will never actually run. Our analysis will err on the side of believing that an application requires a permission that, in fact, it never uses. This means that our estimates of permission bloat are strictly a lower bound on the actual volume of permissions that are requested only to support the needs of the advertising libraries.

Our results, shown in Fig. 2 are quite striking. 15% of apps requesting Internet permissions are doing it for the sole benefit of their advertising libraries. 26% of apps requesting coarse location permissions are doing it for the sole benefit of their advertising libraries. 47% of apps requesting permission to get a list of the tasks running on the phone (the ad libraries use this to check if the application hosting the advertisement is in foreground) are doing it for the sole benefit of their advertising libraries. These results suggest that any architecture that separates advertisements from applications will be able to significantly reduce permission bloat.

## 3 Design objectives

Advertisement services have been around since the very beginnings of the web. Consequently, these services have adapted to use a wide variety of technologies that should be able to influence our AdSplit design.

### 3.1 Advertisement security on the web

Fundamentally, a web page with a third-page advertisement falls under the rubric of a *mashup*, where multiple web servers are involved in the presentation of a single web page.

Many web pages isolate advertisements from content by placing ads in an *iframe* [50]. The content hosted in an iframe is isolated from the hosting webpage and browsers allow only specific cross frame interactions [6, 38], protecting the advertisement against intrusions from the host page (although there have been plenty of attacks [46, 43, 45]). Another valuable property of the iframe is that it allows an external web server to distinguish requests coming from the advertisement from requests that might be otherwise forged. Standard web security mechanisms assist with this; browsers enforce the *same origin policy*, restricting the host web page from making arbitrary connections to the advertiser. Defenses against cross site request forgery, like the Origin header [5], further aid advertisers in detecting fraudulent clicks.

Adapting these ideas to a smartphone requires significant design changes. Most notably, it's common for Android applications to request the privilege to make arbitrary Internet connections. There is nothing equivalent to the same origin policy, and consequently no way for a remote server to have sufficient context, from any given click request it receives, to determine whether that click is legitimate or fraudulent. This requires AdSplit to include several new mechanisms.

## 3.2 Adapting these ideas to AdSplit

The first and most prominent design decision of AdSplit is to separate a host application from its advertisements. This separation has a number of ramifications:

**Specification for advertisements.** Currently the ad libraries are compiled and linked with their corresponding host application. If advertisements are separate, then the host activities must contain the description of of which advertisements to use. We introduced a method by which the host activity can specify the particular ad libraries to be used.

**Permission separation.** AdSplit allows advertisements and host applications to have distinct and independent permission sets.

**Process separation.** AdSplit advertisements run in separate processes, isolated from the host application.

**Lifecycle management.** Advertisements only need to run when the host application is running, otherwise they can be safely killed; similiarly once the host application starts running, the associated advertisement process must also start running. Our system manages the lifecycle of advertisements.

**Screen sharing.** Advertisements are displayed inside host activity, so if advertisements are separated there should be a way to share screen real estate between advertisements and host application. AdSplit includes a mechanism for sharing screen real estate.

**Authenticated user input.** Advertisements generate revenue for their host applications; this revenue is typically dependent on the amount of user interaction with the advertisement. The host application can try to forge user input and generate fraudulent revenue, hence the advertisements should have a way to determine that any input events received from host application are genuine. AdSplit includes a method by which advertising applications can validate user input, validate that they are being displayed on-screen, and pass that verification, in an unforgeable fashion, to their remote server.

In the next section, we will describe how AdSplit achieves these design objectives.

# 4 Implementation

While many aspects of our design should be applicable to any smartphone operating system, we built our system on Android, and there are a number of relevant Android features that are important to describe.

## 4.1 Background

Android applications present themselves to the user as one or more *activities*, which are roughly analogous to windows in a traditional window system. Activities in Android are maintained on a stack, simplifying the user interface and enabling the "back" button to work consistently across applications. This switching between activities as well as other related functions to activity lifecycle are performed by the *ActivityManager* service.

When an activity is started the *ActivityManager* creates appropriate data structures for the activity, schedules the creation of a process for activity, and puts activity-related information on a stack. There is a separate *WindowManager* that manages the z-order of windows and maintains their association with activities. The *ActivityManager* informs the *WindowManager* about changes to activity configuration. Since we want to factor out the advertising code into a separate process / activity, this will require a variety of changes to ensure that the user experience is unchanged.

An app using AdSplit will require the collaboration of three major components: the host activity, the advertisement activity, and the advertisement service. The host activity is the app that the user wants to run, whether a game, a utility, or whatever else. It then "hosts" the advertisement activity, which displays the advertisement. There is a one-to-one mapping between host activity and advertisement activity instances. The Unix processes behind these activities have distinct user-ids and distinct permissions granted to them. To coordinate these two activities, we have a central advertisement service. The ad service is responsible for delivering UI events to the ad activity. It also verifies that the ad activity is being properly displayed and that the UI clicks aren't forged. (More on the verification task in Section 4.4.)

AdSplit builds on QUIRE [13], which prototyped a feature shown in Fig. 3, allowing the host and advertisement activities to share the screen together. First the window for advertisement activity is layered just below the host activity window. The host activity window contains transparent regions where advertisement will be displayed. Standard Android features allow the advertisement activity to verify that the user can actually see the ads.

## 4.2 Advertisement pairing

In AdSplit, we wish to take existing Android applications and separate out their advertising to follow the model described above. We first must explain the variety of different ways in which an existing application might arrange for an advertisement to be displayed. We will use Google's AdMob system as a running example. Other advertisement systems behave similarly, at least with respect to displaying banner ads. (For simplicity of discussion, we ignore full-screen interstitial ads.)

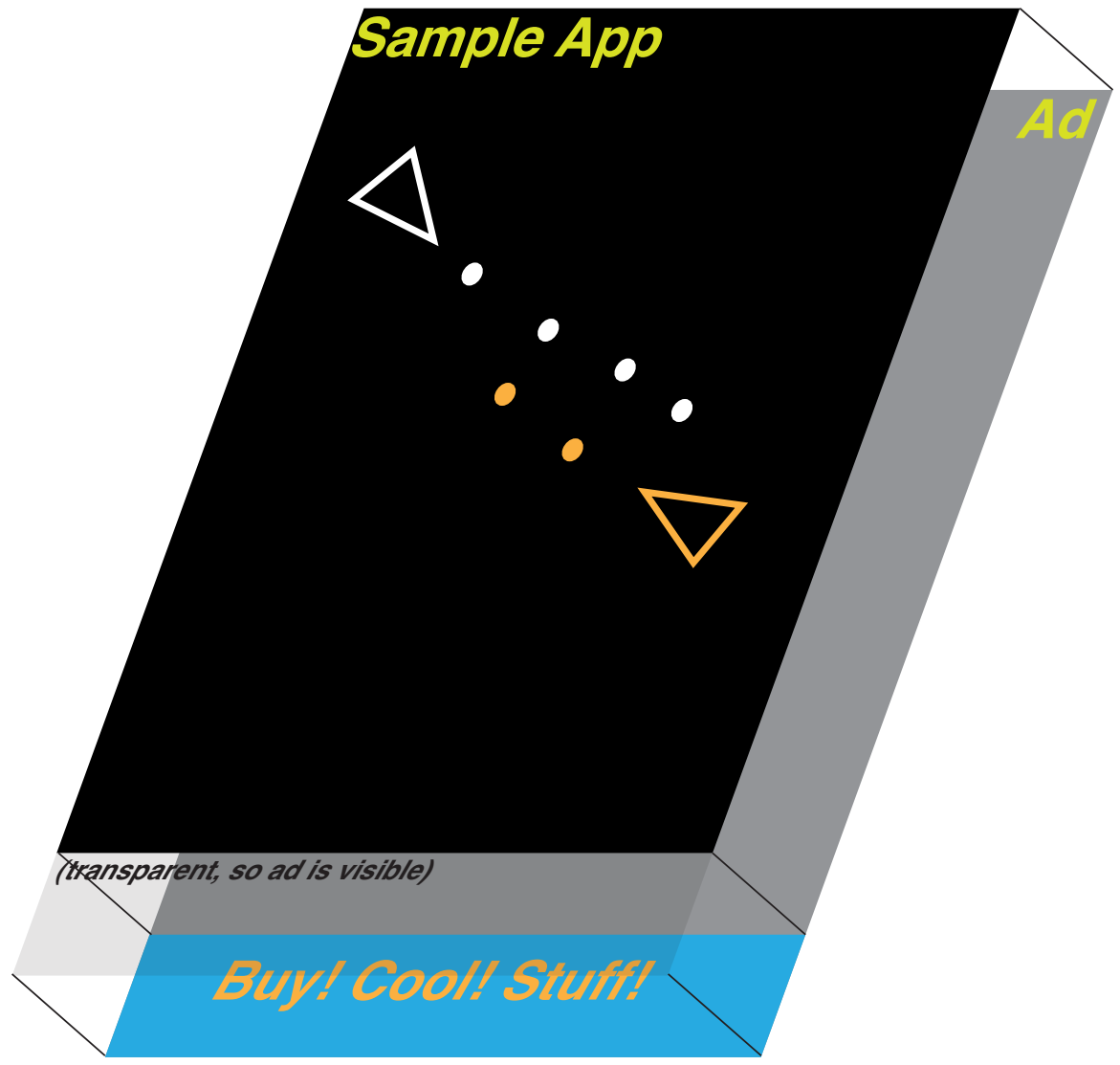

Figure 3: Screen sharing between host and advertisement apps.

With current Android applications, if a developer wants to include an advertisement from AdMob in an activity of her application, she imports the AdMob library, and then either declares an *AdMob.AdView* in the XML layout, or she generates an instance of *AdMob.AdView* and inserts it in directly into the view hierarchy. This works without issue since all AdMob classes are loaded alongside the hosting application; they are separated only by having different package names.

Once we separate advertisements from applications, neither of these techniques will work, since the code isn't there any more. We first need a new mechanism. Later, in Section 6, we will describe how AdSplit does this transformation automatically.

We added a *AppFrame* element, which can appear in the XML manifest, allowing the system to attach a subsidiary activity to its host. This results in a distinct activity for the advertisement as well as a local stub to support the same API as if the advertisement code was still local to the host application. The stub packages up requests and passes them onto the advertisement service.

One complication of this process is that advertising libraries like AdMob were engineered to have one copy running in each process. If we create a single, global instance of any given advertising library, it won't have been engineered to maintain the state of the many original applications which hosted it.

Consequently, the advertisement service must manage distinct advertisement applications for each host application. If ten different applications include AdMob, then there need to be ten different AdMob user-ids in the system, mapping one-to-one with the host applications. The advertisement service is then responsible for ensuring that the proper host application speaks to the proper advertising application.

This is sufficient to ensure that the existing advertising libraries can run without requiring modifications. One complication concerns Android's mechanism for sharing processes across related activities. When a new activity is launched and there is already a process associated with the user-id of the application, Android will launch the new activity in the same process as the old one [2]. If there is already an instance of an activity running, for example, then Android will just resume the activity and bring its activity window to the front of the stack. This is normally a feature, ensuring that there is only a single process at a time for any given application. However, for AdSplit, we need to ensure that advertising apps map one-to-one with hosting apps and we must ensure that their activity windows stay "glued" to their hosts' activities. Consequently, we changed the default Android behavior such that advertisement activities are differentiated based not only by user-id, but also by the host activity. AdSplit thus required modest changes in how activities are launched and resumed as well as how windows are managed.

## 4.3 Permission separation

With Android's install-time permission system, an application requests every permission it needs at the time of its installation. As we described in Section 2, advertising libraries cause significant bloat in the permission requests made by their hosting applications. Our AdSplit architecture allows the advertisements to run as separate Android users with their own isolated permissions. Host applications no longer need to request permissions on behalf of their advertisement libraries.

We note that AdSplit makes no attempt to block a host application from explicitly delegating permissions to its advertisements. For example, the host application might obtain fine-grained location permissions (i.e., GPS coordinates with meter-level accuracy) and pass these coordinates to an advertising library which lacks any location permissions. Plenty of other Android extensions, including TaindDroid [15] and Paranoid Android [42], offer information-flow mechanisms that might be able to forbid this sort of thing if it was considered undesirable. We believe these techniques are complementary to our own, but we note that if we cannot create a hospitable environment for advertisers, they will have no incentive to run in an environment like AdSplit. We discuss this and other policy issues further in Section 8.

## 4.4 Click fraud

AdSplit leverages mechanisms from QUIRE [13] to detect counterfeit events, thus defeating the opportunity for an Android host application to perform a click fraud attack against its advertisers. While a variety of strategies are used to defeat click fraud on the web (see, e.g., Juels et al. [30]), we need distinct mechanisms for AdSplit, since a smartphone is a very different environment from a web browser.

QUIRE uses an system built around HMAC-SHA1 where every process has a shared key with a system service. This allows any process to cheaply compute a "signed statement" and send it anywhere else in the system. The ultimate recipient can then ask the system service to verify the statement. QUIRE uses this on user-generated click events, before they are passed to the host activity. The host activity can then delegate a click or any other UI event, passing it to the advertising activity which will then validate it without being required to trust the host activity. The performance overhead is minimal.

QUIRE has support for making these signed statements meaningful to remote network services. Unlike the web, where we might trust a browser to speak truthfully about the context of an event (see Section 3.1), any app might potentially send any message to any network service. Instead, QUIRE provides a system service that can validate one of these messages, re-sign it using traditional public-key cryptography, and send it to a remote service over the network.

QUIRE's event delivery mechanism is summarized in Fig. 4. The touch event is first signed by the input system and delivered to the host activity. The stub in the host activity then forwards the touch event to advertisement service which verifies the touch event and forwards it to the advertisement activity instance. This could then be passed to another system service (not shown) which would resign and transmit the message as described above.

Despite QUIRE's security mechanisms, there are still several ways the host might attempt to defraud the advertiser. First, a host application might save old click events with valid signatures, potentially replaying them onto an advertisement. We thus include timestamps for advertisements to validate message freshness. Second, host may send genuine click events but move the *AdView* , we prevent this kind of tampering by allowing advertisement service to query layout information about the host activity. Third, a host application might attempt to hide the advertising. Android already includes mechanism for an activity to sort out its visibility to the screen [21] (touch events may include a flag that indicates the window is obscured); our advertising service uses these to ensure that the ad was being displayed at the time the click occurred.

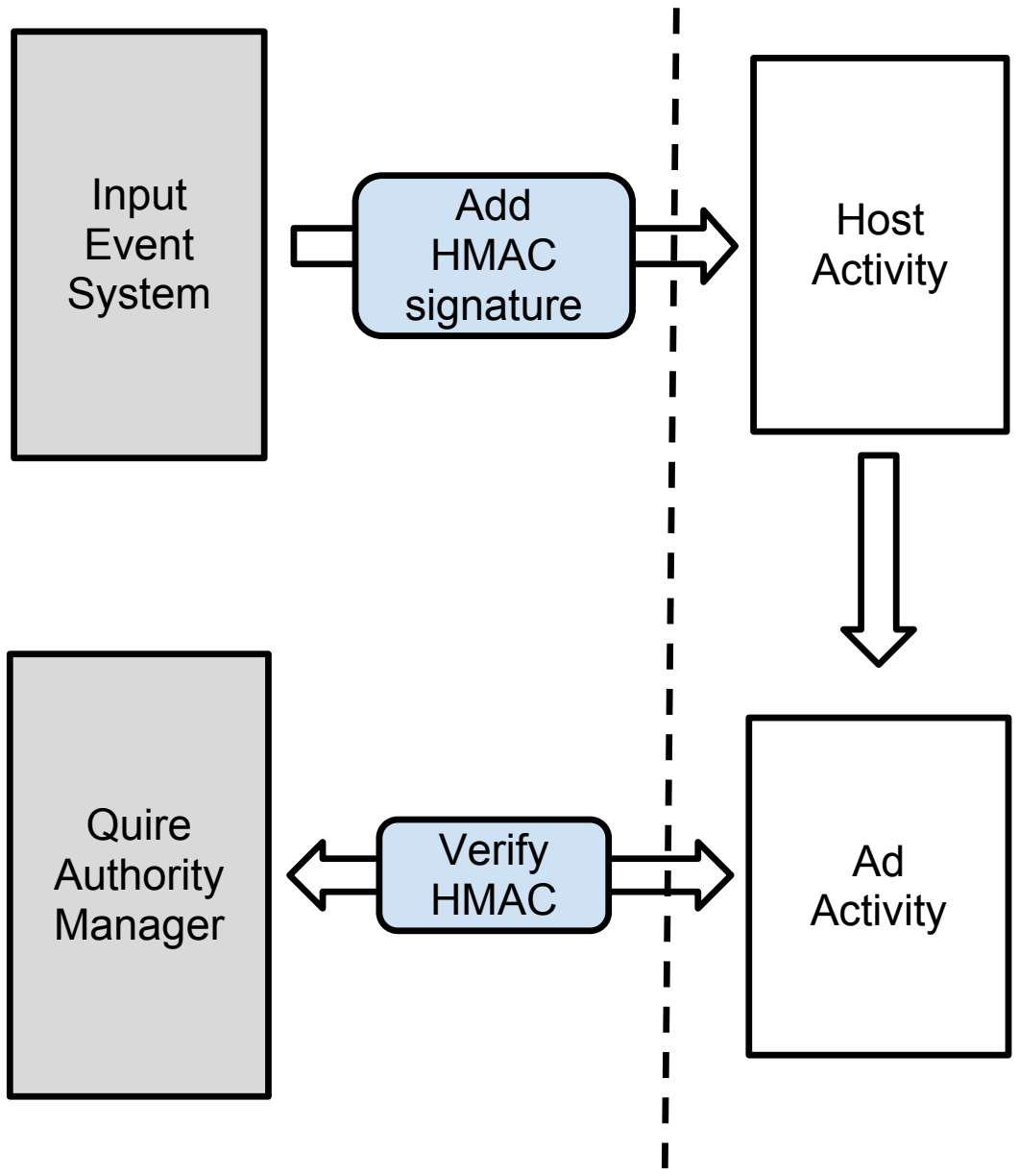

Figure 4: Motion event delivery to the advertisement activity.

It's also conceivable that the host application could simply drop input events rather than passing them to the advertising application. This is not a concern because the host application has no incentive to do this. The host only makes money from clicks that go through, not from clicks that are denied. (Advertising generally works on two different business models: payment per impression and payment per click. In our AdSplit efforts, we're focused on per-click payments, but the same QUIRE authenticated RPC mechanisms could be used in per-impression systems, with the advertisement service making remotely verifiable statements about the state of the screen.)

## 4.5 Summary

AdSplit, as we've described it so far, would not leverage the QUIRE RPC mechanisms by default, since no off-the-shelf advertising library has been engineered to use it. There are other pragmatic issues, such as how the advertisement applications might be installed and managed. We address these issues in Sections 7 and 8. Nonetheless, we now have a workable skeleton design for AdSplit that we have implemented and benchmarked.

## 5 Performance

In order to evaluate the performance overhead of our system we performed our experiments on a standard Android developer phone, the Nexus One, which has a 1GHz ARM core (a Qualcomm QSD 8250), 512MB of RAM, and 512MB of internal Flash storage. We conducted our experiments with the phone displaying the home screen and running the normal set of applications that spawn at start up. We replaced the default "live wallpaper" with a static image to eliminate its background CPU load. All of our benchmarks are measured using the Android Open Source Project's (AOSP) Android 2.3 ("Gingerbread") plus the relevant portions of QUIRE, as discussed earlier.

Our performance analysis focuses on the effect of AdSplit on user interface responsiveness as well as the extra CPU and memory overhead.

### 5.1 Effect on UI responsiveness

We performed benchmarking to determine the overhead of AdSplit on touch event throughput. By default Android has a 60 event per second hard coded limit; for our experiments we removed this limit. Table 2 shows the event throughput in terms of number of touch events per second. (The overhead added by our system is due to passing touch events from the host activity to the advertisement activity. There is also additional overhead due to the additional traversal of the view hierarchy in the advertisement activity.) We can see the our system can still support about 183 events per second which is well above the default limit of 60. Furthermore, the Nexus One is much slower than current-generation Android hardware. CPU overhead, even in this extreme case, appears to be a non-issue.

| **Stock Android** | **AdSplit** | **Ratio** |
|---|---|---|
| 229.96 | 183.12 | 0.796 |

Table 2: Comparison of click throughput (Events/sec), averaged over 1 million events.

### 5.2 Memory and CPU overhead

Measuring memory overhead on Android is complicated since Android optimizes memory usage by sharing read-only data for common libraries. Consequently, if an activity has several copies of a UI widget, the effective overhead of adding a new instance of the same widget is low. Every advertisement library that we examined displays advertisements by embedding a *WebView* . A *WebView* is an instance of web browser. When the host activity already has a *WebView* instance, a fairly common practice, and it includes an advertisement, then most of the code for the advertisement *WebView* will be shared, yielding a relatively low additional overhead for the advertisement. (In our experiments we found out that multiple *WebViews* in the same activity will share their cookies, which means that an advertisement can steal cookies from any other *WebViews* in the activity.)

Consequently, in order to determine the actual memory overhead of separating advertisements from their host applications, we need to differentiate between the cases when host activities contain an instance of *WebView* and when they don't. We did our measurements by running the AdMob library, both inside the application and in a separate advertisement activity. To measure memory overhead we used procrank [14], which tells us the proportional set size (Pss) and unique set size (Uss). Pss is the amount of memory shared with other processes, divided equally among the processes who share it. Uss is the amount of memory used uniquely by the one process. Table 3 lists our results for the memory measurements.

| **Activity setup** | **Memory Overhead (MB)** | | | |
|---|---|---|---|---|
| | **Host Activity** | | **Ad Activity** | |
| | **Pss** | **Uss** | **Pss** | **Uss** |
| Without Ad or WebView | 2.46 | 1.44 | - | - |
| Only WebView | 5.52 | 3.30 | - | - |
| Only AdMob | 9.67 | 6.58 | - | - |
| WebView and AdMob | 9.82 | 6.73 | - | - |
| AdMob with AdSplit | 2.46 | 1.56 | 9.55 | 6.56 |
| WebView and AdMob with AdSplit | 5.15 | 3.35 | 9.29 | 6.58 |

Table 3: Memory overhead for host and advertisement activities with different system configurations.

In interpreting our results we are primarily concerned with the sum of Pss and Uss. From the table, we see that starting with a simply activity without any *WebView* (due to AdMob or its own), consumes about 3.9 MB. This increases to about 9 MB if the activity has a *WebView* . Having AdMob loaded and displaying an advertisement takes about 16.3 MB of memory. When an activity has both *WebView* and AdMob, the total memory used is only about 16.5 MB, demonstrating the efficiency of Android's memory sharing.

With AdMob in a separate process, we expect to pay additional costs for Android to manage two separate ac-tivities, two separate processes, and so forth. The total memory cost in this configuration, with AdMob running in AdSplit and no other *WebView* , is about 20.2MB, roughly a 4 MB increase relative to AdMob running locally. Furthermore, when a separate *WebView* is running in the host activity, there is no longer an opportunity to share the cost of that *WebView* . The total memory use in this scenario is 24.4MB, or roughly an 8 MB increase relative to hosting AdMob locally. We expect we would see similar overheads with other advertising libraries.

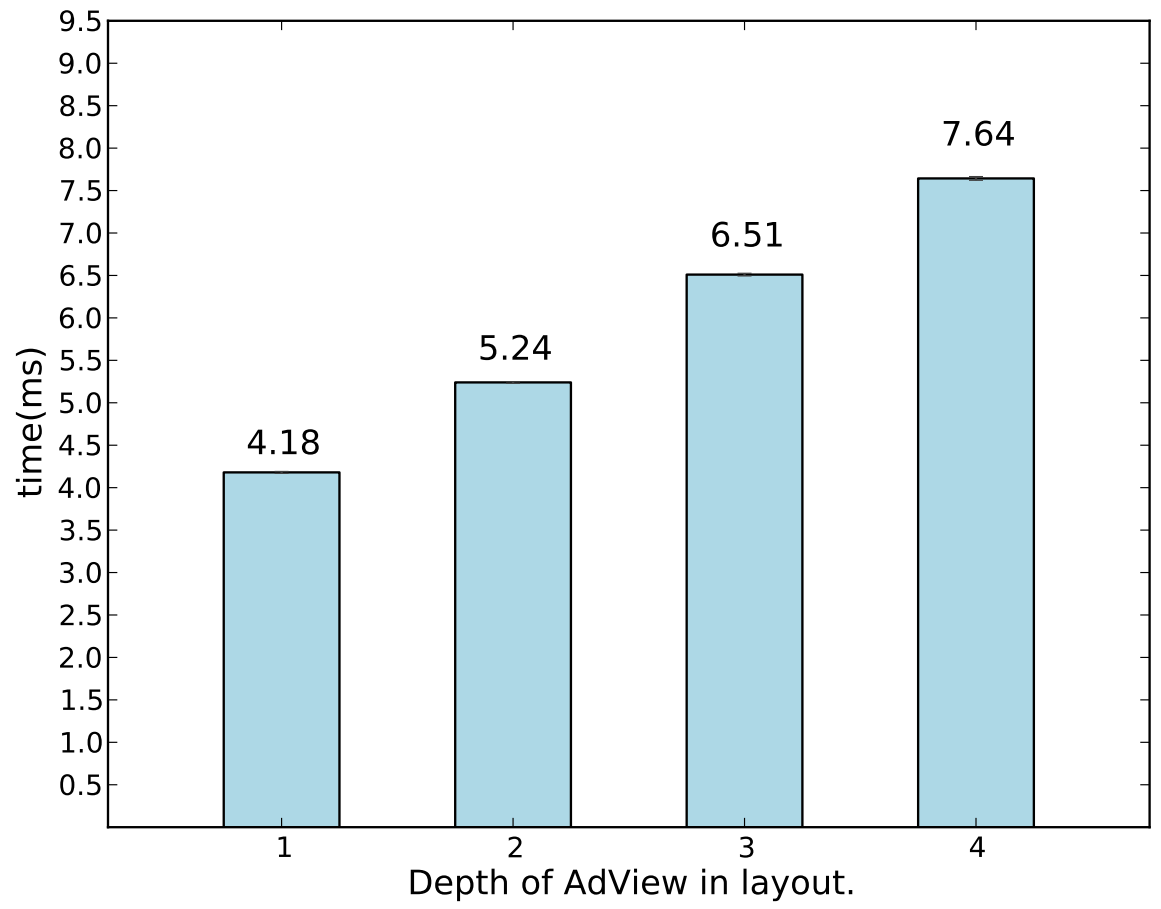

Figure 5: Layout query time vs view depth of host activity (average of 10K runs).

The CPU overhead is same as the overhead of additional Dalvik virtual machine on Android, in fact since the advertisement activities run in background they run with lower priority and can be safely killed without any issues.

As briefly discussed in Section 4.4 we allow advertisement service to query layout information (type, position and transparency of views) about host activity to prevent UI rearrangement attacks. In order to evaluate the overhead of layout information queries we experimented with different view configurations for host activities and varied the depth of *AdView* in the view hierarchy. Fig. 5 shows how the query overhead varies with view depth. The additional depth seems to add about constant (1ms) overhead, which is non-trivial, but we expect these queries to only run once per click.

In summary, while AdSplit does introduce a marginal amount of additional memory and CPU cost, these will have negligible impact in practice.

# 6 Separation for legacy apps

The amount of permissions requested by mobile apps and lack of information about how they are used has been a cause of concern (see, e.g., the U.S. government's Federal Trade Commision study of privacy disclosures for children's smartphone apps [17]). To some extent, the potential for information leakage is driven by advertisement permission bloat, so separating out the ad systems and treating them distinctly is a valuable goal.

As we showed in Section 2, a significant number of current apps with embedded advertising libraries would immediately benefit from AdSplit, reducing the permission bloat necessary to host embedded ads. This section describes a proof-of-concept implementation that can automatically rewrite an Android application to use AdSplit. Something like this could be deployed in an app store or even directly on the smartphone itself.

Figure 6 sketches the rewriting process. First the application is decompiled using android-apktool, converting dex bytecode into smali files. (Smali is to dex bytecode what assembly language is to binary machine code; smali is the human-readable version.) Because smali files are organized into directories based on their package names, it's trivial to distinguish the advertisement libraries from their hosting applications. All we have to do is delete the advertisement code and drop in a stub library, supporting the same API, which calls out to the AdSplit advertisement service. We also analyze the permissions required without the advertisement present (see Section 2) and remove permission requests which are no longer necessary, and edit the manifest appropriately.

For our proof of concept, we decided to focus our attention on AdMob. Our techniques would easily generalize to support other advertising libraries, if desired. (Although we believe we have a better solution, described next in Section 7.)

Our stub library was straightforward to implement. We manually implemented a handful of public methods from the AdMob library, whereafter we constructed a standard Android IPC message to send to the AdSplit advertising service. It worked.

While it would be tempting to use automated tools to translate an entire API in one go, any commercial implementation would require significant testing and, inevitably, there would be corner cases where the automated tool didn't quite do the right thing. Instead, since there are a fairly small number of advertising vendors, we imagine that each one would best be supported by hand-written code, perhaps even supplied directly by the vendor in collaboration with an app store.

Unfortunately, there are a number of significant problems that would stand in the way of an automated rewriting architecture becoming the preferred method of de-ploying AdSplit.

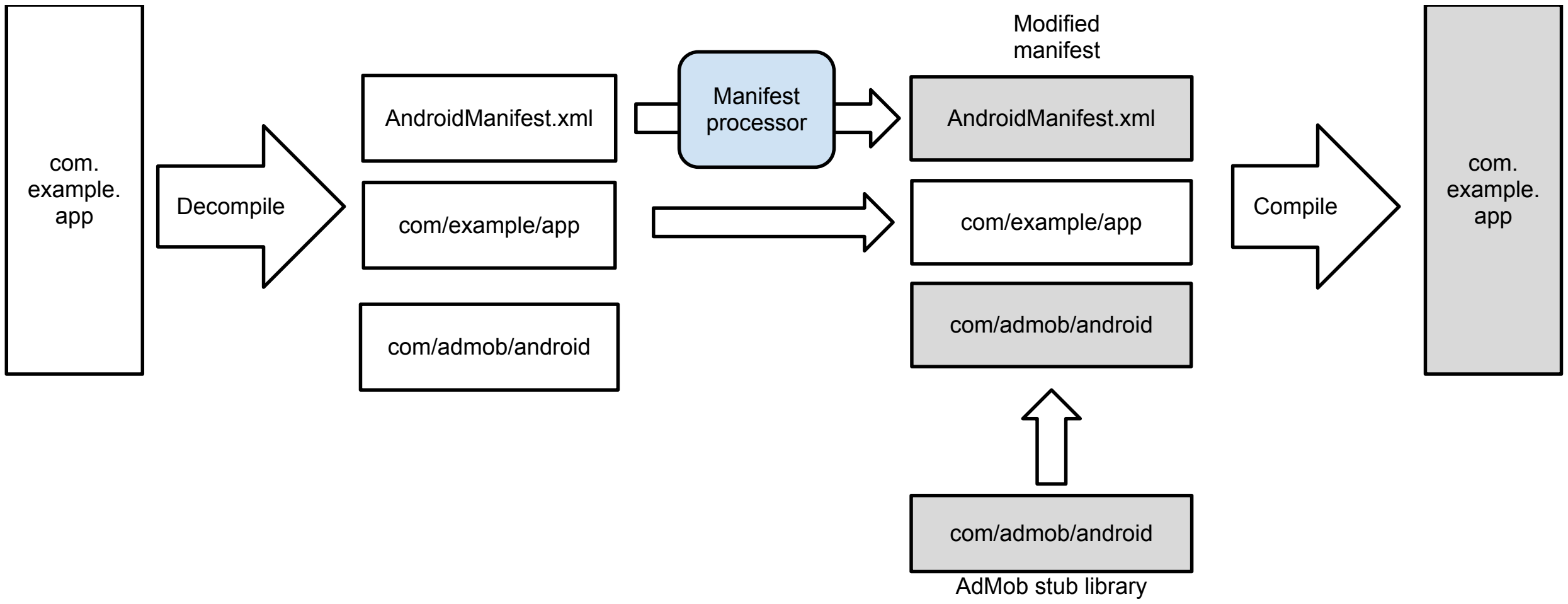

Figure 6: Automated separation of advertisement libraries from their host applications.

**Ad installation.** When advertisements exist as distinct applications in the Android ecosystem, they will need to be installed somehow. We're hesitant to give the host application the necessary privileges to install a third-party advertising application. Perhaps an application could declare that it had a dependency on a third-party app, and the main installer could hide this complexity from the user, in much the same way that common Linux package installers will follow dependencies as part of the installation process for any given target.

**Ad permissions.** Even if we can get the ad libraries installed, we have the challenge of understanding what permissions to grant them. Particularly when many advertising libraries know how to make optional use of a permission, such as measuring the smartphone's location if it's allowed, how should we decide if the advertisement application has those permissions? Must we install multiple instances of the advertising application based on the different subsets of permissions that it might be granted by the host application? Alternatively, should we go with a one-size-fits-all policy akin to the web's same-origin-policy? What's the proper "origin" for an application that was installed from an app store? Unfortunately, there is no good solution here, particularly not without generating complex user interfaces to manage these security policies.

Similarly, what should we do about permissions that many users will find to be sensitive, such as learning their fine-grained location, their phone number, or their address book? Again, the obvious solutions involve creating dialog boxes and/or system settings that users must interact with, which few user will understand, and which advertisers and application authors will all hate.

**Ad unloading.** Like any Android application, an advertisement application must be prepared to be killed at any time—a consequence of Android's resource management system. This could have some destabilizing consequences if the hosting application is trying to communicate with its advertisement and the ad is killed. Also, what happens if a user wants to uninstall an advertising application? Should that be forbidden unless every host application which uses it is also uninstalled?

## 7 Alternative design: HTML ads

While struggling with the shortcomings outlined above with the installation and permissions of advertising applications, we hit upon an alternative approach that uses the same AdSplit architecture. The solution is to expand on something that advertising libraries are already doing: HTML.

If a customer want to purchase advertising on smartphones, they probably want to specify their advertisements the same way they do for the web: as plain text, images, or perhaps as a "rich" ad using JavaScript. Needless to say, a wide variety of tools are available to create and manage such ads, and mobile advertising providers want to make it easy for ads to appear on any platform (iPhone, Android, etc.) without requiring heroic effort from their customer.

Consequently, all of the advertising libraries we examined simply include a WebView within themselves. All of the native Android code is really nothing more than a wrapper around a WebView. Based on this insight, we suggest that AdSplit will be easiest to deploy by providing a single advertising application, build into the Android core distribution, that satisfies the typical needs of Android advertising vendors.

Installation becomes a non-issue, since the only advertiser-provided content in the system is HTML, JavaScript, and/or images. We still use the rest of the AdSplit architecture, running the WebView with a separate user-id, in a separate process and activity, ensuring that a malicious application cannot tamper with the advertisements it hosts. We still have the AdSplit advertisement service, leveraging QUIRE, to validate user events before passing them onto the WebView. We only need to extend the WebView's outbound HTTP transactions to include QUIRE RPC signatures, allowing the remote advertising server to have confidence in the provenance of its advertising clicks.

Security permissions are more straightforward. The same-origin-policy, standard across all the web, applies perfectly to HTML AdSplit. Since the Android WebView is built on the same Webkit browser as the standalone "Browser" application, it has the same security machinery to enforce the same-origin-policy.

Keeping all this in mind we introduced a new form of WebView specifically targetted for HTML ads : the AdWebView. The AdWebView is a way to host HTML ads in a constrained manner. We introduced two advertisement specific permissions which can be controlled by the user. These permissions control whether ads can make internet connections or use HTML5 geoLocation api.

When an ad inside AdWebView requests to load a url or performs call to HTML5 geolocation api, the AdWebView performs a permission check to verify if the associated advertisement origin has the needed advertisement permission. These advertisement permissions can be managed by the user.

About the only open policy question is whether we should allow AdSplit HTML advertisements to maintain long-term tracking cookies or whether we should disable any persistent state. Certainly, persistent cookies are a standard practice for web advertising, so they seem like a reasonable feature to support here a well. AdWebView, by default, doesn't support persistent cookies, but it would be trivial to add.

**Implementation.** We built an advertising application that embeds AdWebView widget, as discussed above. The host application in this case specifies the URL of the advertisement server to be loaded in the AdWebView at initialization. We were successfully in downloading and running advertisements from our sample advertisement server.

**Performance.** Memory and performance overheads are indistinguishable from our AdMob experiments. Both versions host a WebView in a separate process, and it's the same HTML/JavaScript content running inside the WebView.

# 8 Policy

While AdSplit allows for and incentivizes applications to run distinct from their advertisements, there are a variety of policy and user experience issues that we must still address.

## 8.1 Advertisement blocking

Once advertisements run as distinct processes, some fraction of the Android users will see this as an opportunity to block advertisements for good. Certainly, with web browsers, extension like AdBlock and AdBlock Plus are incredibly popular. The Chrome web store lists these two extensions in its top six[1] with "over a million" installs of each. (Google doesn't disclose exact numbers.)

The Firefox add-ons page offers more details, claiming that AdBlock Plus is far and away the most popular Firefox extension, having been installed just over 14 million times, versus 7 million for the next most popular extension[2]. The Mozilla Foundation estimates that 85% of their users have installed an extension [37]. Many will install an ad blocker.

To pick one example, Ars Technica, a web site popular with tech-savvy users, estimated that about 40% of its users ran ad blockers [33]. At point, it added code to display blank pages to these users in an attempt to cajole them into either paying for ad-free "premium" service, or at least configuring their ad blocker to "white list" the Ars Technica website.

Strategies such as this are perilous. Some users, faced with a broken web site, will simply stop visiting it rather than trying to sort out why it's broken. Of course, many web sites instead employ a variety of technical tricks to get around ad blockers, ensuring their ads will still be displayed.

Given what's happening on the web, it's reasonable to expect a similar fraction of smartphone users might want an ad blocker if it was available, with the concomitant arms race in ad block versus ad display technologies.

So long as users have not "rooted" their phones, a variety of core Android services can be relied upon by host applications to ensure that the ads they're trying to host are being properly displayed with the appropriate advertisement content. Similarly, advertising applications (or

[1] https://chrome.google.com/webstore/category/popular

[2] https://addons.mozilla.org/en-US/firefox/extensions/?sort=users

HTML ads) can make SSL connections to their remote servers, and even embed the proper remote server's public key certificate, to ensure they are downloading data from the proper source, rather than empty images from a transparent proxy.

Once a user has rooted their phone, of course, all bets are off. While it's hard to measure the total number of rooted Android phones, the CyanogenMod Android distribution, which requires a rooted phone for installation, is installed on roughly 722 thousand phones[3]—a tiny fraction of the hundreds of millions of Android phones reported to be in circulation [41]. Given the relatively small market share where such hacks might be possible, advertisers might be willing to cede this fraction of the market rather than do battle against it.

Consequently, for the bulk of the smartphone marketplace, *advertising apps on Android phones offer greater potential for blocking-detection and blocking-resistance than advertising on the web*, regardless of whether they are served by in-process libraries or by AdSplit. Given all the other benefits of AdSplit, we believe advertisers and application vendors would prefer AdSplit over the status quo.

## 8.2 Permissions and privacy

Some advertisers would appear to love their ability to learn additional data about the user, including location information, address book, other apps running on the phone, and so forth. This information can help profile a user, which can help target ads. Targeted ads, in turn, are worth more money to the advertiser and thus worth more money to the hosting application. When we offer HTML style advertisements, with HTML-like security restrictions, the elegance of the solution seems to go against the higher value profiling that advertisers desire.

Leaving aside whether it's *legal* for advertisers to collect this information, we have suggested that a host application could make its own requests that violate the users' privacy and pass these into the AdSplit advertising app. We hope that, if we can successfully reduce apps' default requests for privileges that they don't really need, then users will be less accustomed to seeing such permission requests. When they do occur, users will push back, refusing to install the app. (Reading through the user-authored comments in the Android Market, many apps with seemingly excessive permission requirements will often have scathing comments from users, along with technical justifications posted by the app authors to explain why each permission is necessary.)

Furthermore, if advertisers ultimately prefer the AdSplit architecture, perhaps due to its improved resistance to click fraud and so forth, then they will be forced to make the trade-off between whether they prefer improved integrity of their advertising platform, or whether they instead want less integrity but more privacy-violating user details.

[3] http://stats.cyanogenmod.com/

# 9 Related Work

## 9.1 Web security

AdSplit considers an architecture to allow for controlled mashups of advertisements and applications on a smartphone. The web has been doing this for a while (as discussed in Section 3.1). Additionally, researchers have considered a variety of web extensions to further contain browser components in separate processes [25, 44], including constructing browser-based multi-principal operating systems [27, 49].

## 9.2 JavaScript sandboxes

Caja [35] and ADsafe [1] work as JavaScript sandboxes which use static and dynamic checks to safely host JavaScript code. They use a safe subset of JavaScript, eliminating dangerous primitives like `eval` or `document.write` that could allow an advertisement to take over an entire web page. Instead, advertisements are given a limited API to accomplish what they need. AdSplit can trivially host advertisements built against these systems, and as their APIs evolve, they could be directly supported by out AdWebView class. Additionally, because we run the AdWebView in a distinct process with its own user-id and permissions, we provide a strong barrier against advertisement misbehavior impacting the rest of the platform.

## 9.3 Advertisement privacy

Privad [26] and Juels et al. [29] address security issues related to privacy and targeted advertising for web ads. They use client side software that prevents behavior profiling of users and allows targeted advertisements without compromising user privacy.

AdSplit does not address privacy problems related to targeted advertisements but it provides framework for implementing various policies on advertisements.

## 9.4 Smart phone platform security

As mobile phone hardware and software increase in complexity the security of the code running on a mobile devices has become a major concern.

The Kirin system [16] and Security-by-Contract [12] focus on enforcing install time application permissions within the Android OS and .NET framework respectively. These approaches to mobile phone security allow a user to protect themselves by enforcing blanket restrictions on what applications may be installed or what installed applications may do, but do little to protect the user from applications that collaborate to leak data or protect applications from one another.

Saint [40] extends the functionality of the Kirin system to allow for runtime inspection of the full system permission state before launching a given application. Apex [39] presents another solution for the same problem where the user is responsible for defining run-time constraints on top of the existing Android permission system. Both of these approaches allow users to specify static policies to shield themselves from malicious applications, but don't allow apps to make dynamic policy decisions.

CRePE [11] presents a solution that attempts to artificially restrict an application's permissions based on environmental constraints such as location, noise, and time-of-day. While CRePE considers contextual information to apply dynamic policy decisions, it does not attempt to address privilege escalation attacks.

#### 9.4.1 Privilege escalation

XManDroid [8] presents a solution for privilege escalation and collusion by restricting communication at runtime between applications where the communication could open a path leading to dangerous information flows based on Chinese Wall-style policies [7] (e.g., forbidding communication between an application with GPS privileges and an application with Internet access). While this does protect against some privilege escalation attacks, and allows for enforcing a more flexible range of policies, applications may launch denial of service attacks on other applications (e.g., connecting to an application and thus preventing it from using its full set of permissions) and it does not allow the flexibility for an application to regain privileges which they lost due to communicating with other applications.

One feature of Quire that is not used in AdSplit is its ability to defeat confused deputy attacks, by annotating IPCs with the entire call chain. In concurrent work to Quire, Felt et al. present a solution to what they term "permission re-delegation" attacks against deputies on the Android system [20]. With their "IPC inspection" system, apps that receive IPC requests are poly-instantiated based on the privileges of their callers, ensuring that the callee has no greater privileges than the caller. IPC inspection addresses the same confused deputy attack as Quire's "security passing" IPC annotations, however the approaches differ in how intentional deputies are handled. With IPC inspection, the OS strictly ensures that callees have reduced privileges. They have no mechanism for a callee to deliberately offer a safe interface to an otherwise dangerous primitive. Unlike Quire, however, IPC inspection doesn't require apps to be recompiled or any other modifications to be made to how apps make IPC requests.

(AdSplit does not require Quire's IPC inspection system, and thus also does not require apps to be recompiled to have the semantics described in this paper.)

More recent work has focused on kernel extensions that can observe IPC traffic, label files, and enforce a variety of policies [9, 47]. These systems can enhance the assurance of many of the above techniques by centralizing the policy specification and enforcement mechanisms.

#### 9.4.2 Dynamic taint analysis on Android

The TaintDroid [15] and ParanoidAndroid [42] projects present dynamic taint analysis techniques to preventing runtime attacks and data leakage. These projects attempt to tag objects with metadata in order to track information flow and enable policies based on the path that data has taken through the system. TaintDroid's approach to information flow control is to restrict the transmission of tainted data to a remote server by monitoring the outbound network connections made from the device and disallowing tainted data to flow along the outbound channels.

AdSplit allows ads to run in separate processes but applications can still pass sensitive information to separated advertisements. TaintDroid and ParanoidAndroid can be used to detect and prevent any such flow of information. Thus they are complementary to AdSplit.

## 10 Future Work

The work in this paper touches on a trend that will become increasingly prevalent over the next several years: the merger of the HTML security model and the smartphone application security model. Today, HTML is rapidly evolving from its one-size-fits-all security origins to allow additional permissions, such access to location information, for specific pages that are granted those permissions by the user. HTML extensions are similarly granted varying permissions rather than having all-or-nothing access [4, 32].

On the flip side, iOS apps originally ran with full, unrestricted access to the platform, subject only to vague policies enforced by human auditors. Only access to location information was restricted. In contrast, the Android security model restricts the permissions of apps, with many popular apps running without any optional permissions at all. Despite this, Android malware is a growing problem, particularly from third-party app stores (see, e.g., [19, 51]). Clearly, there's a need for more restrictive Android security, more like the one-size-fits-all web security model.

While it's unclear exactly how web apps and smartphone apps will eventually become one thing, our paper shows where this merger is already underway: when web content is embedded in a smartphone app. Well beyond advertising, a variety of smartphone apps take the strategy of using native code to set up one or more web views, then to the rest in HTML and JavaScript. This has several advantages: it makes it easier to support an app across many different smartphone platforms. It also allows authors to quickly update their apps, without needing to go through a third-party review process.

These trends, plus the increasing functionality in HTML5, suggest that "native" apps may well be entirely supplanted by some sort of "mobile HTML" variant, not unlike HP/Palm's WebOS, where every app is built this way[4].

Maybe this will result in a industry battle royale, but it will also offer the ability to ask a variety of interesting security questions. For example, consider the proposed "web intents" standard[5]. How can an "external" web intent interact safely with the "internal" Android intent system? Both serve essential the same purpose and use similar mechanisms. We, and others, will pursue these toward interesting conclusions.

## 11 Conclusion

We have presented AdSplit, an Android-based advertising system that provides advertisers integrity guarantees against potentially hostile applications that might host them. AdSplit leverages several mechanisms from Quire to ensure that UI events are correct and to communicate to the outside world in a fashion that hosting applications cannot forge. AdSplit runs with marginal performance overhead and, with our HTML-based design, offers a clear path toward widespread adoption. AdSplit not only protects advertisers against click fraud and ad blocking, it also reduces the need for permission bloat among advertising-supported free applications, and has the potential to reduce the incentive for applications to leak privacy-sensitive user information in return for better advertising revenues.

## Acknowledgements

This work builds on our prior Quire project. We thank Yuliy Pisetsky and Anhei Shu for their assistance and efforts. This work was supported in part by NSF grants CNS-1117943 and CNS-0524211.

[4] http://developer.palm.com/blog

[5] http://webintents.org/